\documentclass[prl,amsmath,amssymb,twocolumn,times]{revtex4-1}

\usepackage{amsmath,amssymb}
\usepackage{graphicx}
\usepackage[usenames]{color}
\usepackage{amssymb}
\usepackage{grffile}

\newcommand{\beq}{\begin{eqnarray}}
\newcommand{\eeq}{\end{eqnarray}}

\def \av#1{{\langle#1\rangle}}
\def \ket#1{{|#1\rangle}}

\newcommand{\nin}{{n_{\textrm{in}}}}

\def \figwa{0.49}

\def \figw{0.36}
\def \figw3{0.24}

\newcommand{\cpht}{Centre de Physique Th\'eorique, \'Ecole Polytechnique, CNRS,
91128 Palaiseau, France}
\newcommand{\unige}{D\'epartement de Physique Th\'eorique, Universit\'e de Gen\`eve,
1211 Gen\`eve, Switzerland}

\begin{document}

\title{Controllable manipulation and detection of local densities and bipartite entanglement in a quantum gas by a dissipative defect}

\author{Peter Barmettler}
\author{Corinna Kollath}

\affiliation{\unige}
\affiliation{\cpht}

\date{\today}

\begin{abstract}
We study the complex dynamics of a one-dimensional Bose gas subjected to a dissipative local defect which induces one-body atom losses. In experiments these atom losses occur for example when a focused electron or light beam or a single trapped ion is brought into contact with a quantum gas. We discuss how within such setups one can measure or manipulate densities locally and specify the excitations that are induced by the defect. In certain situations the defect can be used to generate entanglement in a controlled way despite its dissipative nature. The careful examination of the interplay between hole excitations and the collapse of the wave function due to non-detection of loss is crucial for the understanding of the dynamics we observe.
\end{abstract}

\maketitle

Ultracold quantum gases offer the possibility to address many open questions from various areas of physics due to their excellent control and fast tunability. Achievements of the last decade range from the observation of strongly correlated one-dimensional bosonic gases, so-called Tonks-Girardeau gases~\cite{Paredes2004Kinoshita2004}, to the realization of a Mott-insulating state in optical lattices, both for bosonic~\cite{Greiner2002a} and fermionic~\cite{Jordens2008Schneider2008} atoms. Only very recently, new experimental setups opened the way to resolve single atoms in quantum gases. Fluorescence techniques \cite{Bakr2010Sherson2010,Weitenberg2011Zimmermann2011} or scanning with a highly focused electron beam \cite{Gericke2008} can address single columns in the quantum gas with a resolution of the optical lattice spacing. Real three-dimensional resolution could be reached using a trapped ion \cite{Zipkes2010Schmid2010,Kollath2007Sherkunov2009}.
These local techniques can be employed to detect or even to manipulate quantum gases. Consequently, these tools enable one to take advantage of the presence of the trap to access a large range of the homogeneous phase diagram \cite{Ho2009Nascimbene2010} in a single realization of an experiment.

The main subject of this work is to explore how a highly localized dissipative defect that generates atom losses, such as an ion or a focused electron or light beam, can be used in a one-dimensional system to probe and to manipulate properties of the quantum state. For instance, we show how to measure the local density efficiently and explain how shock waves emerge in superfluid states. At strong dissipative strength subsystems left and right from the defect separate spatially but entanglement between the subsystems can be manipulated in a controlled way and is directly related to the experimentally measurable atom loss. Our theoretical model is a correlated one-dimensional bosonic lattice system which is coupled to a Markovian environment by one-body atom losses. Actually, the situation of a local defect immersed in a quantum system is very general and not only of interest in the context of cold atomic gases but also in condensed matter physics. It represents a non-trivial many-body problem and in order to capture the coaction of continuously created losses and intrinsic dynamics of the correlated quantum states correctly, the full treatment of the master equation of the interacting system beyond mean-field \cite{Brazhnyi2009,Shchesnovich2010} is crucial \cite{Shchesnovich2010a}.  

In a broad parameter regime, a one-dimensional Bose-gas which is subjected to an optical lattice potential can be well described by the Bose-Hubbard model, 
\begin{eqnarray}
H=\sum_{\ell} -J\left( b^\dagger_{\ell}b_{\ell+1}+b^\dagger_{\ell+1}b_{\ell}\right)+ \frac{U}{2}n_\ell(n_\ell-1).
	\label{eq:bhmodel}
\end{eqnarray}
Here $b^\dagger_{\ell}$ ($b_\ell$) are the bosonic creation (annihilation) operators at site $\ell$ and $n_\ell$ is the density operator. We assume a system with open boundaries; integer indices $\ell$  ranging from $-(L-1)/2$ to $(L-1)/2$ represent individual sites that are spaced by distance $a$. The first term models the kinetic energy of the atoms in the periodic potential and the second term the on-site interaction due to s-wave scattering. 
Initially, in the absence of a dissipative defect, the atomic cloud is prepared in its ground state. For weak interaction this ground state is a superfluid with large on-site density fluctuations. At integer filling the state is Mott-insulating for interactions above a critical strength (in 1D $(U/J)_c\approx 3.4$ at filling $n=1$). In contrast, at non-integer filling the weakly interacting superfluid is connected by a crossover to the strongly interacting Tonks-Girardeau gas of impenetrable bosons.

At time $t=0$ the dissipative defect (localized at site $\ell=0$), e.g. a trapped ion or an electron or light beam, is brought into contact with the bosonic cloud and starts generating atom losses at the central lattice site only. Experiments have reached or are close to reach such resolution \cite{Gericke2008,Bakr2010Sherson2010,Zipkes2010Schmid2010}. From the theoretical point of view it is of secondary importance if one or few sites are affected. We assume that atoms are expelled quickly (compared to the timescales of the quantum gas) to the free space continuum and that the defect itself does not become correlated to the quantum gas. For this case a Markovian approximation can be used to derive a master equation \cite{Carmichael1993GardinerBook} for the density matrix $\rho$ of the bosonic cloud,
	$\dot{\rho}=\mathcal{L^C}\rho+\mathcal{L^D}\rho,\quad
	\mathcal{L^C}\rho := -i[H,\rho],$  with the dissipator
	$\mathcal{L^D}\rho:=\Gamma b_0\rho b^\dagger_0-\frac{\Gamma}{2}n_0\rho- \frac{\Gamma}{2}\rho n_0$.
The strength of the coupling to the environment $\Gamma$ depends on the cross-section of the scattering processes between the atoms and the defect.
In the ion-atom hybrid system the loss mechanism can be seen as an elastic scattering event of an atom with a hot ion. In principle, its strength can be widely tuned by using a suitable magnetic Feshbach resonance \cite{Idziaszek2009} or using different kinds of atoms and ions.
Typical energies of an ion are of the order of $1k_B$K, while the atomic trapping potential is of the order of $1k_B\mu$K \cite{Zipkes2010Schmid2010}. Therefore the atoms will be scattered to the free continuum in a time much shorter than the many-body timescales in the cloud (e.g. $J\sim 10^{-8}k_B$K).  When shining a focused electron beam on the cold atomic sample \cite{Gericke2008} the loss mechanism corresponds to an ionization process, for which a similar separation of energy scales is present (in this case $\Gamma$ can be tuned by the intensity of the beam). Also in experiments with fluorescent light \cite{Weitenberg2011Zimmermann2011} the energy gain by spontaneous emission is sufficient to transfer an atom to the continuum or at least to higher bands. Contrarily to previous experiments \cite{Weitenberg2011Zimmermann2011} in which the lattice was increased prior to illumination in order to avoid this process, in the current proposal the lattice is kept constant and the focus is on the induced  dynamics.

The non-equilibrium problem of the master equation is solved by sampling over quantum trajectories \cite{Carmichael1993GardinerBook,Molmer1993}. 
The propagation of the stochastic wave functions with the non-hermitian Hamiltonian $H_{\text{eff}}=H-i\Gamma n_0$ \cite{Molmer1993} is achieved by the time-dependent density matrix renormalization group (DMRG) \cite{Daley2004White2004,Daley2009}. The imaginary part of $H_{\text{eff}}$ represents the effect of non-detection of atom loss. Detection of atom loss is simulated by annihilation of bosonic fields  at stochastically chosen times. For the study of the dynamics of locally created defects in one-dimensional systems, DMRG works quite efficiently as seen in a spin system in Ref. \cite{Gammelmark2010}. In the present context, bond dimensions 100-200 in the DMRG and time-steps of $0.001-0.005\, \hbar/J$ lead to errors negligible compared to the statistical error of the sampling procedure. The optimal choice of DMRG parameters is important, since averaging over more than $6000$ trajectories may be necessary. Quantum trajectories allow for a natural definition of entanglement entropy in this open system \cite{Hyunchul2004}. One defines $S_l$, the entropy of entanglement for a bipartition of the system at a site $l$, by the average over the von Neumann entanglement entropies of single trajectories.

One of the experimentally accessible quantities is the total atomic loss $N(t)=\sum_\ell\left[n_\ell(t)-n_\ell(t=0)\right]$ measured after a given time $t$  ($n_\ell(t)=\av{n_\ell}(t)$). 
On short time scales, as long as current flows into the central site are negligible (guaranteed for $t\ll\hbar/J$, $t\ll\hbar/\Gamma$), the total loss is 
\begin{equation}
	N(t)=\nin  (1-e^{-\Gamma t/\hbar})\mbox{, where } \nin=n_0(t=0).
	\label{eq:st}
\end{equation}
We will focus on the response of the system for times beyond this short time behaviour in which a complex many-body dynamics sets in that depends on the nature of the underlying quantum state in a non-obvious manner.

\begin{figure}[tb]
	\begin{center}
		\includegraphics[width=0.48\textwidth]{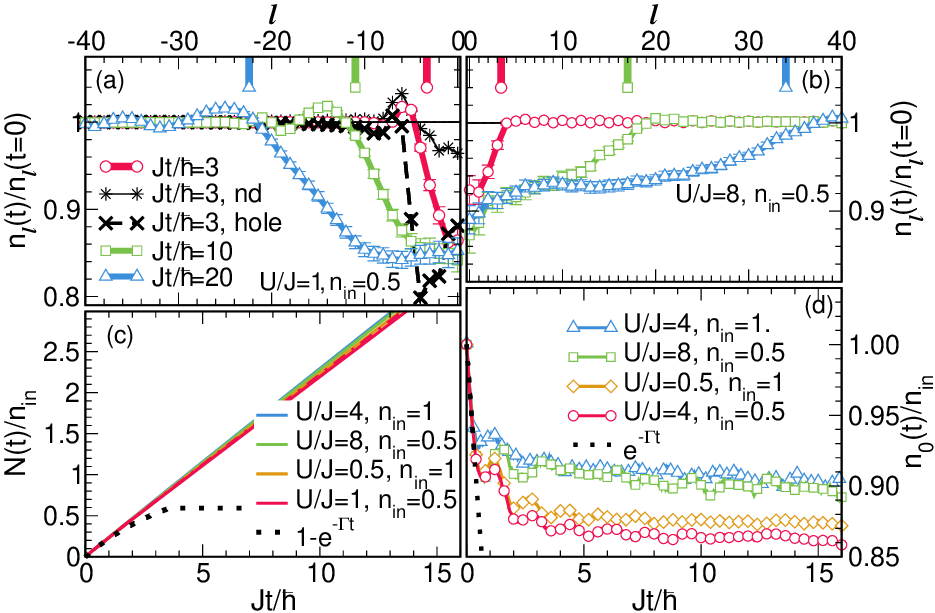}
		\caption{\label{fig:Nn1} (Color online) Weak dissipative coupling $\Gamma/J=0.25$, $L=105$ and 53 or 105 particles.  In (a) and (b) we show density profiles at different times. Densities are rescaled by their initial values and size-dependent features in the vicinity of the edges of the systems are discarded. We depict the 'light-cone' of waves moving with the sound velocity \cite{Kollath2005} by vertical lines. Star and cross symbols represent $\ket{\psi_\text{nd}(t)}$, $\ket{\psi_\text{hole}(t)}$ respectively. In the lower panels we plot the total atom loss (c) and the central density (d), both rescaled by the initial density $\nin $. In (c) 4 curves almost lie on top of each other. Boundary effects are eliminated by interrupting simulations before reaching the recurrence times. Statistical errors are either marked by bars or smaller than line width or symbol size.}
\end{center}
\end{figure}

First we analyse the time evolution of the density profiles in the case of a weak dissipative defect, $\Gamma/J=0.25$ [Fig. \ref{fig:Nn1}(a-b)]. For the strongly interacting Tonks gas [$U/J=8$, $\nin=0.5$ in Fig.~\ref{fig:Nn1}(b)]  a dark density modulation moving at the speed of sound (close to $2Ja /\hbar$ \cite{Kollath2005}) is the dominating feature in the relaxation. Nearly identical evolutions of the density were found for a Mott insulating state (not shown), in which hole modes \cite{Huber2007} are also propagating with velocity $v\approx2Ja/\hbar$. The insulating nature does not play an important role for the transport of the excitation. In contrast to these strongly interacting cases, in the weakly interacting superfluid state [$U/J=1$, $\nin=0.5$ in Fig.~\ref{fig:Nn1}(a)] we observe the formation of a remarkably strong bright shock wave at a speed which strongly exceeds the sound velocity.
Only after this density wave the expected dark modulation propagates.

To reveal the origin of this shock wave we analyze in Fig.~\ref{fig:Nn1}(a) density profiles of single trajectories which allow to separate the effects of the detection and non-detection of atom losses: (i) the wave function evolves in time by the Hamiltonian and the non-detection part of the dissipative action only, $\ket{\psi_\text{nd}(t)}=e^{-iH_{\text{eff}}t/\hbar}\ket{\psi_0}$; (ii) the wave function is the evolution of a hole, generated at $t=0$ by a single local atom loss, with the Bose-Hubbard Hamiltoninan (\ref{eq:bhmodel}), $\ket{\psi_{\text{hole}}(t)}=e^{-iHt/\hbar}b_0\ket{\psi_0}$. From this analysis follows clearly that the continuously induced collapse of the wave function by the non-detection of atom losses is the driving force behind the shock waves. The propagation of a localized hole does not lead to significant shock wave formation although in the present case the dark perturbation in the density of $\ket{\psi_{\text{hole}}(t)}$ is much larger than in $\ket{\psi_\text{nd}(t)}$. The highly non-linear effect of the non-detection of loss is directly related to the non-local correlations present in the superfluid and therefore not observed in the more strongly interacting cases.

\begin{figure}[ht]
	\begin{center}
		\includegraphics[width=0.48\textwidth]{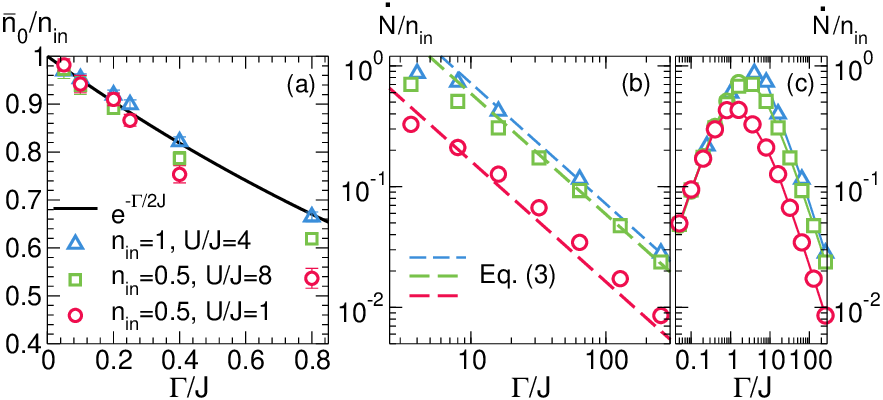}
	\end{center}
	\caption{(Color online) (a) $\overline{n}_0$ in the weakly dissipative regime. Error bars indicate systematic fluctuations and statistical errors. (b) Long-time loss rate for the strongly dissipative case. (c) Loss rate from weak to strong dissipation. Lines are guides to the eye. \label{fig:largetime}
}
\end{figure}

Surprisingly almost no sign of this very different dynamics due to the underlying quantum phase can be noticed in the evolution of the total atom losses [Fig. \ref{fig:Nn1}(c)]. At short times, before a current sets in, $N(t)$ should obey the exponential law (\ref{eq:st}) for any type of system, which is indeed the case up to $t\approx\hbar/4J$. More interestingly, at larger times $N(t)/\nin$ shows a linear increase with practically no dependence on the initial state. In  Fig.~\ref{fig:Nn1}(d) the central density $n_0(t)$  is plotted in order to visualize the tiny differences in the behaviour of the total loss (it represents the loss rate by the relation $\dot{N}(t)=\Gamma n_0(t)$). $n_0(t)$ describes how, through a slowly decaying, oscillatory transient regime, a quasi-steady value is reached. In order to investigate systematically the onset of current we extract a stationary value $\overline{n}_0$ by averaging over the time interval $10\hbar/J<t<15\hbar/J$.  Looking at $\overline{n}_0$ for various dissipative coupling strength [Fig.~\ref{fig:largetime}(a)] we found a generically obeyed relation $\overline{n}_0/\nin \sim e^{-\Gamma/2J}$ up to coupling strength $\Gamma/J\approx0.4$. From this result we can conclude that $\hbar/2J$ is the characteristic timescale for the onset of a current. From our analysis of the density profiles we can understand this time scale from the modes that dominate the relaxation process. Indeed, from the velocities of sound and 'hole' modes observed in Tonks and Mott systems ($v\approx 2Ja/\hbar$) a relaxation time $\hbar/2J$ would be expected.  In principle, for the weakly interacting superfluid one has a much smaller sound velocity \cite{Kollath2005} ($v\approx 1.1 Ja/\hbar$ in the case plotted in Fig. \ref{fig:Nn1}(a)). However, we found that the faster shock-waves are the dominant modes, which apparently again lead to a relaxation time scale $\hbar/2J$. 

Hence, the short-time exponential decay (\ref{eq:st}) followed by a long time linear rise with slope $\sim e^{\Gamma/2J}$ is generic and obeyed for very different types of systems. This means that one can actually infer the value of the initial density from atom loss measurements without explicit knowledge of the nature of the state of the sample beyond the short time regime (\ref{eq:st}). Long exposure of the sample to the defect is helpful to increase the detected signal.

\begin{figure}[ht]
	\begin{center}
		\includegraphics[width=\figwa\textwidth]{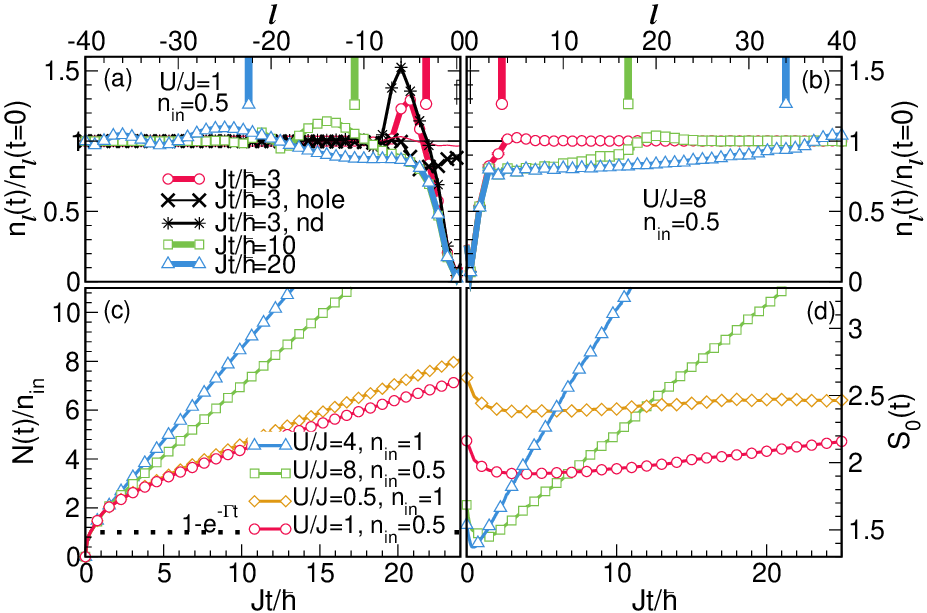}
	\end{center}
\caption{\label{fig:Nn2} (Color online) Numerical results for strongly dissipative coupling, $\Gamma/J=8$. For explanation of (a-c)  see Fig. \ref{fig:Nn1}. Panel (d) shows the evolution of the von-Neuman entropy between the subsystems separated by the defect.}
\end{figure}

Let us turn to the response of the atomic cloud to a strong dissipative defect ($\Gamma\gg J,U$). The density profiles [Fig.~\ref{fig:Nn2}(a-b)] show the propagation of modes that are very similar to the weakly dissipative case, but with more pronounced density modulations. At the central site the strong dissipation leads to a very rapid suppression of the density suggesting a decoupling of the system at the defect. In contrast to the previously considered weakly dissipative case, the atom loss is now highly sensitive on the initial state [Fig.~\ref{fig:Nn2}(c)]. The von Neumann entropy $S_0$ also strongly depends  on the initial state and shows quite unexpected behaviour: strong entanglement between the two separated subsystems can be generated by the defect [Fig.~\ref{fig:Nn2}(d)]. Only at short times the entropy of entanglement is reduced by the projection performed by the measurement. At larger times the entanglement grows approximately linearly in time. In the strongly interacting systems the linear rise is found numerically to be proportional to the atom loss $\dot{S}_0(t)\approx \alpha \dot{N}(t)$, $\alpha\leq 0.5$, where a maximal efficiency $\alpha=0.5$ is reached in the hard-core limit $U/J\rightarrow \infty$ (not shown). For the curves at strong but finite interactions plotted in Fig. ~\ref{fig:Nn2}(d) we find $\alpha\approx 0.3$. The effect of non-detection of atom loss is crucial for this specific relation between loss and entanglement entropy. If there was only creation of holes at different instances of times we would expect that each hole, which splits into two mutually entangled excitations propagating into opposite directions, would lead to an increase in the entanglement entropy by one unit. However, the collapse of the wave function by non-detection of loss destroys such pairs and apparently limits the efficiency of entanglement generation to $\alpha\leq 0.5$. 
When decreasing the interaction, efficiency $\alpha$ is lowered and the production of entanglement becomes nearly completely suppressed in the weakly interacting superfluids [Fig. ~\ref{fig:Nn2}(d)]. This could be explained by the presence of density fluctuations in the initial wave function. The propagating hole pairs do not create new highly excited states, as it is the case at strong interactions, but only leads to redistribution of weights between states.

It is interesting to gain an analytical understanding of the long time-limit of the density profiles and the corresponding atom loss, especially because of their relation to the entanglement entropy in strongly interacting states.  A perturbative analysis by adiabatic elimination of occupations of the central site (see e.g.~\cite{Garcia-Ripoll2009}) can be performed. From the effective master equation so derived,  $\!\!\dot N(t\!\gg \!\!\hbar/J)\! =\!\frac{2J^2}{\Gamma} \!\!\!\sum_{s=\pm1}  \!\!\!\left[ 2n_{s}(t)\!+\! \av{b^\dagger_sb_{-s}}(t)\right] \!\!+O\left(\frac{J^3}{\Gamma^2}\right)\!,$
it follows that the loss rate decreases at increasing coupling strength as $\frac{J^2}{\Gamma}$. This is a counterintuitive result \cite{Gammelmark2010,Shchesnovich2010}, which is known for discrete measurements as the quantum Zeno effect \cite{Misra1977}.  Using that the density close to the center can be described by a static boundary situation and correlations between the two sites next to the defect can be neglected, the perturbative expression can be further simplified,
\begin{equation}
	\dot{N}(t\gg \hbar/J) \approx \frac{8J^2}{\Gamma} n_{\pm (L-1)/2}(t=0).\label{eq:pert2}
 \end{equation}
 A comparison between this expression and the numerical results gives an excellent agreement down to a dissipative coupling of $\Gamma\approx8J$ for strong interactions [Fig.~\ref{fig:largetime}(b)] (the loss rate is extracted by averaging over the stationary state for $10\hbar/J<t<15\hbar/J$.).
In contrast, the agreement to the weakly interacting superfluid is not yet fully established. The averaged values nevertheless obey the characteristic $J^2/\Gamma$ scaling with a strong tendency towards the stationary rate. We expect that for longer times than were accessible to us, this value will be reached. 

In conclusion, we have shown that dissipative coupling can be used to measure densities accurately and to manipulate properties of a quantum gas in a controlled way, although it induces highly non-trivial many-body dynamics. Rather unexpected is the strong growth of entanglement between seemingly decoupled subsystems. It is very interesting for practical applications that this entanglement can be manipulated, and, at the same time, be monitored by the experimentally easily accessible atom loss. We have calculated these effects in a one-dimensional system with great precision, but the main features, e.g.~density measurements or the Zeno-effect, are expected to occur in higher dimensions as well. Further, entanglement generation should be possible analogously when placing a one-dimensional defect into a two-dimensional lattice. 
In experiments, one interesting avenue to follow now would be to use a moving defect to measure static or time-resolved correlation functions or to manipulate different regions of the system. Moreover, the loss mechanism can be used to remove entropy-rich regions  and thus may serve as a tool to further cool down an atomic gas. The new experimental tools that manipulate quantum gases by local dissipation open a new perspective on many-body dynamics of correlated systems.

\emph{Acknowledgments} -- We acknowledge discussions with the group of M. K\"ohl, 
J.-S. Bernier, C. Berthod, J. Dalibard, V. Guarrera, M. Heyl, H. Ott, D. Poletti and support by SNF, ANR (FAMOUS), DARPA-OLE and 'Triangle de la 
Physique'.

\end{document}